\documentclass[pra,showpacs,twocolumn,nofootinbib]{revtex4}
\usepackage{graphicx}
\setkeys{Gin}{draft=false} 
\usepackage{bm,amsmath}
\usepackage{dcolumn}
\usepackage{natbib}
\usepackage
{hyperref}

\newcommand{\fref}[1]{Fig.~\ref{#1}}

\newcommand{\Eref}[1]{Eq.~(\ref{#1})}

\def\veps{\varepsilon}

\begin{document}

\title{Exchange assisted tunneling and positron annihilation on inner atomic shells}

\author{M. G. Kozlov$^{1,2,3}$}
\author{V. V. Flambaum$^{1,4}$}

\affiliation{$^1$School of Physics, The University of New South Wales, Sydney
NSW 2052, Australia}

\affiliation{$^2$Petersburg Nuclear Physics Institute, Gatchina 188300,
             Russia}
\affiliation{$^3$St.~Petersburg Electrotechnical University ``LETI'', Prof.
Popov Str. 5, 197376 St.~Petersburg}

\affiliation{$^4$Centre for Theoretical Chemistry and Physics, New Zealand
Institute for Advanced Study, Massey University, Auckland 0745, New Zealand}

\date{
\today}

\begin{abstract}
It is known for a long time that the long range asymptotic behavior of the
Hartree-Fock orbitals is different from that of the orbitals in the local
potential. However, there is no consensus about observable physical effects
associated with this asymptotics. Here we argue that weaker decrease of the
Hartree-Fock orbitals at large distances is responsible for the positron
annihilation on the inner shell electrons, which is observed experimentally.
\end{abstract}

\pacs{
 31.15.A-,     
 03.65.Ge      
 } \maketitle

\section*{Long distance behaviour of electron orbitals}

It is sometimes assumed that exchange interaction between electrons is
important only at short distances and can be neglected when one of the
electrons is far from the origin. It is easy to see that this is not true if
we consider long range asymptotics of an inner shell electron
\cite{HMS69,DFS82,Fla09c}. At large distances the exchange term for all
occupied orbitals has the form $r^{-\nu} \phi_v$, where $\phi_v$ is the
outermost orbital with the highest energy $\veps_v$. The power $\nu$ depends
on the leading multipolarity of the exchange interaction. The monopole
component does not contribute to the asymptotics due to the orthogonality of
the orbitals and for the dipole component of the exchange interaction $\nu=2$.

This fact was first realized by \citet{HMS69}, who showed that generally
speaking asymptotic behavior of all Hartree-Fock orbitals is given by the
exponential with the highest energy $\veps_v$ (or the smallest binding
energy):
\begin{align}\label{eq_HF}
  \phi_i^\mathrm{hf}|_{r\gg r_v}
  \sim r^{-\nu_i}\exp{\left(-\sqrt{-2\veps_v}r\right)}\,,
\end{align}
where $\nu_i$ is specific for each orbital and $r_v$ is the radius of the
outermost atomic orbital. We use atomic units $\hbar=m_e=|e|=1$ unless stated
otherwise. Because the monopole component of the exchange interaction does not
contribute to the asymptotic behaviour, the expression \eqref{eq_HF} does not
apply to the systems where only $s$-orbitals are occupied.

Later the asymptotic behaviour of the electron orbitals was reanalyzed by many
authors \cite{MPL75,KaDa80,DFS82,Fla09b,Fla09c}. \citet{DFS82} showed that the
asymptote \eqref{eq_HF} holds for the relativistic Hatree-Fock-Dirac equations
as well. The role of correlations were studied by \citet{MPL75} and
\citet{KaDa80}, who demonstrated that \Eref{eq_HF} holds also for the natural
orbitals $\phi_i^\mathrm{nat}$ with nonzero occupation numbers. Natural
orbitals are the eigenfunctions of the one electron density matrix
$\rho(x',x)$, where $x\equiv \bf{r},\sigma$ and they can be found for any many
electron wavefunction.

Natural orbitals are mostly used in the context of the configuration
interaction approach. For the core-valence correlations the many-body
perturbation theory (MBPT) is usually more efficient. Within MBPT approach
\citet{Fla09c} showed that \Eref{eq_HF} also holds for the Brueckber orbitals.
The latter are solutions of the one particle equation with the correlation
potential $\Sigma$ added to the Hartree-Fock potential. The nonlocal part of
the potential $\Sigma$ decreases faster than the exchange potential and
therefore does not change the long range behaviour of the Brueckner orbitals.

In the solid state physics the long range exchange induced interaction is well
known. For example, the Ruderman-Kittel-Kasuya-Yasida (RKKY) exchange-induced
spin-spin interaction is responsible for the magnetic ordering in metal alloys
\cite{RuKi54,Kas56,Yos57}. The long range interaction in one- two- and
three-dimensional systems was also considered in \cite{Fla09c}.

In spite of the results cited above and numerous other studies it is sometimes
assumed that asymptotic behavior \eqref{eq_HF} is an artifact of the used
approximations and physically observable one particle asymptote should depend
on the energy $\veps_i$ of the orbital in question (see, for example, Ref.\
\cite{FMP98}):
\begin{align}\label{eq_spec}
  \phi_i|_{r\gg r_v}
  \sim r^{-\nu_i}\exp{\left(-\sqrt{-2\veps_i}r\right)}\,.
\end{align}
Such asymptotic behavior follows, for example, from the analytical
continuation to the negative energies of the wave function in the scattering
theory \cite{KaDa80}:
\begin{align}\label{eq_scatt}
 \Psi^N &= -\frac{S_{0,i}}{r_1}\mathrm{e}^{i(k_ir_1+\eta_i\log r_1)}
 \times\Psi^{N-1}_i(2\dots N)\,,
\end{align}
where $S_{0,n}$ is the matrix element of the $S$ matrix and $\Psi^{N-1}_i$ is
the wavefunction of the ion. However it is easy to present an example where
\Eref{eq_spec} is incorrect: the double well potential with one electron below
and another above the barrier (see Appendix \ref{sec_DW} and also Ref.\
\cite{Fla09b}).

To get asymptote \eqref{eq_spec} it is sufficient to assume that when the
first electron is far from the origin, $r_1\gg 1$, the $N$ particle
wavefunction $\Psi^N$ can be factorized as
\begin{align}\label{eq_fact1}
  \Psi^N|_{r_1\gg 1} &=\phi_i(1)\times\Psi^{N-1}_i(2\dots N)\,.
\end{align}
The $N$-particle Hamiltonian can be written in a form:
 \begin{multline}\label{eq_fact2}
 h(1)+h(2)\dots +h(N) + \sum_{k<l}\frac{1}{r_{kl}}
 \\ = h(1) + H^{N-1}(2\dots N) + \sum_{l=2}^N \frac{1}{r_{1l}}
 \\ \approx h(1) + H^{N-1}(2\dots N) + \sum_{l=2}^N
 \left(\frac{1}{r_{1}}+\frac{r_l}{r_{1l}^2}\right)\,.
 \end{multline}
In the last line we left two first terms in the expansion of $1/r_{1l}$ and
skipped for simplicity the angular factors $P_1(\cos \theta_{1l})$.
Substituting Eqs.\ \eqref{eq_fact1} and \eqref{eq_fact2} in the Schr\"odinger
equation $H^N\Psi^N = E^N\Psi^N$ and integrating over the coordinates $2\dots
N$ with the function $\Psi^{*\,N-1}_i$ we get:
 \begin{align}\label{eq_fact3}
 \left(h(r_1) + \frac{N-1}{r_1}\right)\phi_i(r_1)
 &= \left(E^N-E^{N-1}_i\right)\phi_i(r_1)\,.
 \end{align}
Due to the parity selection rule the second term of the multipolar expansion
from \eqref{eq_fact2} turns to zero here. The solution of this equation
satisfies \Eref{eq_spec} with the energy
 \begin{align}\label{eq_eps}
 \veps_i=E^N-E^{N-1}_i\,.
 \end{align}

The obvious problem with the anzatz \eqref{eq_fact1} is that it is not
antisymmetric in permutations. It also does not account for the correlations
between the outgoing electron and the electrons of the remaining ion. Let us
see what happens if we substitute \eqref{eq_fact1} with the wave function of
the general form $\Psi^N$. Following \cite{KaDa80} we can project $N$-particle
wave function on the eigenstates of the $(N-1)$-particle ion:
 \begin{subequations}
 \label{ap45}
 \begin{align}
 &|\Psi^{N}(1\dots N)\rangle
 =\sum_i f_i(1) |\Psi_i^{N-1}(2\dots N)\rangle\,,
 \label{ap4}
  \\
 &f_i(1) = \langle\Psi_i^{N-1}(2\dots N)|\Psi^{N}(1\dots N)
 \rangle_{2\dots N}\,.
 \label{ap5}
 \end{align}
 \end{subequations}
Expansion \eqref{ap4} is valid at all distances and functions $f_i$ play the
role of one-particle orbitals. They do not form an orthonormal basis set. In
general these functions are: (i) not orthogonal to each other; (ii) normalized
so that $\sum_i \langle f_i|f_i\rangle=1$; (iii) not described by a definite
angular momentum $j_i$. It follows from \Eref{ap4} that each function $f_i$
has several angular components:
 \begin{align}
 f_i(\bm{r}) = \sum_{j=|J-J_i|}^{J+J_i} f_{i,j,m}(\bm{r})
  \,,
 \label{ap6a}
 \end{align}
where $m=M-M_i$ and $J,M$ and $J_i,M_i$ are the angular quantum numbers of the
initial atom and the final ion respectively.

It is easy to see that the amplitudes $f_i$ satisfy following equations
\cite{KaDa80}:
 \begin{subequations}
 \label{ap78}
 \begin{multline}
 \label{ap7}
 \left(h(\bm{r}_1) - \veps_i\right)f_i(\bm{r}_1)
 \\= -\sum_{k} f_k(\bm{r}_1) W_{i,k}(\bm{r}_1)\,,
 \end{multline}
 \begin{multline}
 \label{ap8}
 W_{i,k}(\bm{r}_1)
 = \langle\Psi^{N-1}_i(2\dots N)|\sum_{l=2}^N
 \frac{1}{r_{1,l}}|\Psi^{N-1}_k(2\dots N)\rangle
 \\
 =\int{\frac{\rho^{N-1}_{i,k}(\bm{r}_2,\bm{r}_2)}{r_{1,2}}
 \mathrm{d} \bm{r}_2}
 \,,
 \end{multline}
 \end{subequations}
where $\rho^{N-1}_{i,k}(\bm{r}',\bm{r})$ is the transition density matrix of
the ion.

If we look for the solution of this system at large distances it is useful to
single out from the functions $W_{i,i}$ the term with zero multipolarity and
add it to the left hand side:
 \begin{multline}
 \label{ap9}
 \left(h(\bm{r}_1) + \frac{N-1}{r_1}- \veps_i\right)f_i(\bm{r}_1)
 \\ = \left((N-1)\frac{f_i(\bm{r}_1)}{r_1}
 -\sum_{k} f_k(\bm{r}_1) W_{i,k}(\bm{r}_1)\right)\,.
 \end{multline}
This way we account for the screening of the nucleus by the electrons of the
ion (note that this expression is still exact). Now we can take a limit
$r_1\gg 1$ and apply multipolar expansion to evaluate the integral in
\Eref{ap8}. The zeroth multipole cancels the first term in the right hand side
of \Eref{ap9} and we get
 \begin{multline}
 \label{ap9a}
 \left(h({r}_1) + \frac{N-1}{r_1}- \veps_i\right)f_i({r}_1)
 \\ = \frac{1}{r_1^2}\sum_k{}\!' f_k(r_1)
 \int{\rho^{N-1}_{i,k}(r,r)r\,r^2\mathrm{d} r}\,.
 \end{multline}
Here we again omitted the angular factors and assumed that the sum runs only
over states $k$ which satisfy dipole selection rule.

Let us assume that the solution of the homogeneous equation is localized at
the distances $r_i$. Then, for the larger distances $r_1\gg r_i$ we can
neglect the solution of the homogeneous equation \eqref{eq_fact3} and write:
 \begin{multline}\label{eq_fact4a}
 f_i(r_1)|_{r_1\gg r_i}
 = \left(h(r_1) + \frac{N-1}{r_1}-\veps_i\right)^{-1}
 \\
  \frac{1}{r_1^2}\sum_k{}\!' f_k(r_1)
 \int{\rho^{N-1}_{i,k}(r,r)r^3\mathrm{d} r}
 \,.
 \end{multline}
This expression can be simplified further if we note that at large distances
the resolvent is approaching a constant \cite{Fla09c}, $\left(h(r) +
\tfrac{N-1}{r}-\veps_i\right)^{-1} \rightarrow -\veps_i^{-1}$, so
 \begin{align}\label{eq_fact4b}
 f_i(r)|_{r\gg r_i}
 \approx -
 \frac{1}{\veps_i r^2}\sum_k{}\!' f_k(r)\!
 \int{\rho^{N-1}_{i,k}(y,y)y^3\mathrm{d} y}
 \,.
 \end{align}

In Eqs.\ (\ref{ap9a} -- \ref{eq_fact4b}) we did not make any assumptions about
the wavefunction of the ion. In the single determinant approximation the
integral in \eqref{eq_fact4b} is reduced to the sum over occupied orbitals of
the opposite parity, $p_i p_k=-1$ \cite{DFS82}:
 \begin{align}\label{eq_fact5}
 f_i(r)|_{r\gg r_i}
 \approx -\frac{1}{\veps_i r^2}
  \sum_k{}\!' f_k(r)\langle \phi_k|r|\phi_i\rangle
 \,,
 \end{align}
and we return to the Hartree-Fock case.

Both equations \eqref{eq_fact4b} and \eqref{eq_fact5} clearly lead to the
asymptote \eqref{eq_HF}. For $r\gtrsim r_v$ the outermost orbital dominates
the sum and
 \begin{align}\label{eq_fact5a}
 f_i(r)|_{r\gtrsim r_v}
 \sim r^{-\nu_i}f_v(r)
 \,.
 \end{align}
Note that the power $\nu_i$ can be larger than 2 if there is no dipole matrix
element $\langle f_v|r|f_i\rangle$. In this case higher terms of the
multipolar expansion are required. Expression \eqref{eq_eps} shows that
instead of the one electron binding energy of the valence orbital $-\veps_v$
one should use the ionization potential of the atom $I^{N}$:
\begin{align}\label{eq_gen}
  f_i|_{r\gg r_v}
  \sim r^{-\nu_i}\exp{\left(-\sqrt{2I^{N}}r\right)}\,.
\end{align}

We see that \Eref{eq_gen} has much wider applicability than the Hartree-Fock
approximation. On the other hand it is valid only for the distances $r\gg
r_v$, where for the neutral atom $r_v\sim 1$. In fact, for the inner orbitals
the exchange interaction starts to dominate over the direct Coulomb
interaction much earlier, at $r\gg r_i$, where $r_i$ is the radius of the
inner orbital in question \cite{DFS82,Fla09c,Fla09a}. The more general form of
\Eref{eq_gen} is given by \Eref{eq_fact4b}. In particular we can use them to
estimate inner orbitals in the classically forbidden region. At large
distances in the many electron atom all inner orbitals are asymptotically
proportional to the outermost orbital up to a power of the radius. For shorter
distances different terms of the sum \eqref{eq_fact5} will dominate.

Note that the system \eqref{ap78} is particularly useful for the highly
charged ions, where $\tfrac{1}{Z}$ expansion is applicable. In this case the
right hand side in \Eref{ap7} is of the order $\tfrac{1}{Z}$ and we can solve
this system iteratively. For the first order corrections to the amplitudes
$f_i$ we can use zero order wavefunctions $\Psi^{N-1\,(0)}_i$ to calculate
functions $W_{i,k}$. This way we can find analytical form of the first order
corrections in $\tfrac{1}{Z}$ to the amplitudes $f_i$. One simple example is
considered in Appendix \ref{sec_He}.

In the end of this section we can say that asymptotic behaviour of the inner
orbitals is changed by the entanglement induced by the (anti) symmetrization
postulate. This effect appears in the Hartree-Fock approximation, but survives
for the correlated many-electron atoms.

\section*{Observability of the one-particle asymptotics of the bound many-electron
wave function}

Here we try to formalize the notion of the one-particle asymptotics of the
bound many-electron wave function. In quantum mechanics one need to associate
an operator with any observable. When we discuss asymptotic behavior of the
bound orbitals we mean that electron is registered at a given distance $R$
from the origin and we simultaneously register the ion in a state with the
hole in a given shell. This can be described by the following operator:
 \begin{subequations}
 \label{ap12}
 \begin{align}
 T_{i}^{R} &= \sum_{l=1}^N T_{l,i}^{R}\,,
 \label{ap1}
 \\
 T_{1,i}^{R} &= |\Psi_i^{N-1} (2\dots N)\rangle
 \frac{\delta(r_1\!-\!R)}{4\pi R^2}
 \langle\Psi_i^{N-1} (2\dots N)|\,,
 \label{ap2}
 \end{align}
 \end{subequations}
and so on for the operators $T_{2,i}$, $T_{3,i}$, etc. When this operator is
applied to the antisymmetric wave function $\Psi^N$, we get:
 \begin{align}
 \langle\Psi^{N}|T_{i}^{R}|\Psi^{N}\rangle
  &=
  N\langle\Psi^{N}|T_{1,i}^{R}|\Psi^{N}\rangle\,.
 \label{ap3}
 \end{align}
Expectation value of the operator $T^R_i$ for state \eqref{ap4} is:
 \begin{align}
 \langle\Psi^{N}|T_{i}^{R}|\Psi^{N}\rangle
  &= \frac{N}{4\pi R^2}
  \langle f_i(\bm{r})|\delta(r\!-\!R)|f_i(\bm{r})\rangle
  \,.
 \label{ap6}
 \end{align}
Thus, we can say that functions $f_i$ indeed play the role of the orbital,
whose long-range behavior we want to study.

The observable \eqref{ap12} does not commute with the Hamiltonian. The
measurement of $T_i$ requires significant energy. In order to detect an
electron at a particular distance $R$ we introduce uncertainty in its
momentum. That means that we interact with the whole many-electron system.
Because of that the energy of the system is not conserved and we detect final
ion with the energies $E_i^{N-1}$, which are larger than the initial energy of
the system $E^N$. The energy we need to detect position of the bound electron
is $E\gg\veps_i$.

We see that the measurement of the observable $T_i$ changes the energy of the
system. This is different from the observations in the scattering theory where
we detect particles at infinity. Such measurements can be done with arbitrary
small momentum and energy transfer. Because of that in the scattering theory
the energy of the system is conserved. These examples suggest that when during
the measurement the energy of the system is conserved the exponent of the
asymptote is given by the energies of the initial and the final state. On the
contrary, when the measurement requires energy, the asymptote can be described
by many exponents.

\section*{Positron annihilation on inner electrons}

It seems that asymptotics \eqref{eq_spec} and \eqref{eq_gen} are so different
that it should be easy to prove  experimentally which of them is correct.
However these expressions coincide for the outermost electron shell, which
gives dominant contribution to any physical processes at large distances. The
inner shell asymptotic amplitudes are strongly suppressed. Though predictions
for the inner shell contribution to such processes from Eqs.\ \eqref{eq_spec}
and \eqref{eq_gen} can differ by many orders of magnitude, they still can be
too small to be experimentally observable.

Recently \citet{Amu09} argued that asymptotics (\ref{eq_gen}) should lead to
an observable field ionization (FI) from the inner shells in the strong
electric field ${\bf F}= F\hat{\bf z}$. He estimated that the inner shell
contribution scales as $F^4$ and for the field $F\sim 1\,\mathrm{a.u.} =
5\times 10^9$ V/cm it can be on the order of $10^{-5}$. Unfortunately such
static fields are not achievable. If we use the low frequency laser field
instead of the dc field, it will be hardly possible to disentangle tunneling
FI from the multi photon processes. Therefore, it is unlikely that FI from the
inner shells can be detected even for the asymptotics given by \Eref{eq_gen}.

Here we want to draw attention to the positron annihilation on the inner
electron shells \cite{IGGS97,EGKK02,DuGr06,WSGS10} as a potential test of the
expression (\ref{eq_fact5a}). In the non-relativistic approximation the
annihilation vertex is proportional to the $\delta$-function. Thus, if we can
control the final state of the ion, the annihilation cross section can be
approximately linked to the observable $T(R)$ defined by \Eref{ap12}.

In the typical experiment only one of the two annihilation gamma quanta is
detected (see, e.g. the review \cite{SGB05}). For a low energy positrons an
observed linewidth is determined by a Doppler broadening associated with the
average momentum of the bound electron. Because of that the annihilation
linewidth is given by \cite{DuGr06}:
 \begin{align}\label{eq_width}
 \Gamma\approx \sqrt{|\veps_i| m_e c^2}\approx 3.7\times\sqrt{|\veps_i|}\,
 \mathrm{KeV}\,.
 \end{align}
Thus, the annihilation on the inner shells contributes to the wings of the
line, while annihilation on the outermost shell gives the central peak.
Consequently the accurate study of the annihilation line shape provides
information about the inner shell contribution \cite{IGGS97}. The inner shell
annihilation was also detected directly with the coincidence technique for
$\gamma$ quanta and Auger electrons \cite{EGKK02}.

Let us discuss \Eref{eq_width} in more detail. The line shape of the
annihilation $\gamma$ line is determined by the Fourier transform of the
atomic orbital. It is clear that inner orbitals have wider spectrum and
contribute to the wings of the line. However it is not so clear that the
asymptotic part of the inner orbitals also contributes to the wings. Let us
write the inner orbital as
\begin{align}\label{eq_a1}
  \phi_i=\phi_i^0+\phi_i^a\,,
\end{align}
where $\phi_i^0$ is a solution in a local potential and $\phi_i^a$ describes
exchange induced asymptotic tail. Let us write $\phi_i^a$ as
\begin{align}\label{eq_a2}
  \phi_i^a(r)=\frac{C}{r^2+a^2}\phi_v(r)\,.
\end{align}
This function has correct asymptotic behaviour \eqref{eq_fact5a} with
$\nu_i=2$ and parameter $a$ is introduced to insure that at short distances
$\phi_i^0 \gg \phi_i^a$. The Fourier transform of this function is the
convolution of the transforms of the multipliers. The width of the convolution
is the sum of the widths of the components, i.e. $P_i^a \approx a^{-1}+P_v$,
where $P_v\approx r_v^{-1}$ is the width of the Fourier transform of the
valence orbital $\phi_v$. The exchange term starts to dominate the tail at the
distances between $r_i$ and $r_v$ \cite{DFS82}. Thus, the cutoff parameter $a$
should be chosen from the interval $r_i< a < r_v$. Even if we take $a$ close
to $r_v$ we get the width two times larger than for the outermost orbital.
However, $a$ is closer to $r_i$ since it plays a role of the cut-off factor in
the expansion over $r_</r_>$ in the exchange Coulomb interaction between inner
and outer electron. For $a \sim r_i$ we return to \Eref{eq_width}. In both
cases the annihilation on the asymptotic part of the inner orbital $\phi_i^a$
indeed contribute to the wing of the $\gamma$ line.

We conclude that in the annihilation experiment we can control the final state
of the ion. Unfortunately it is not the case for the position of the
annihilated electron. Still, the positron can not penetrate deep into the atom
because of the strong repulsion from the nucleus. Therefore, the annihilation
should predominantly take place at the edge of the atom.

Let us estimate classical turning point for the thermal positron with the
energy $\veps_p\sim 300\,\mathrm{K} \sim 10^{-3}$ a.u. Atomic potential seen
by the slow positron has the form \cite{DFKM93}:
\begin{align}\label{eq_U}
  U(r)=Q(r)/r-\alpha/2r^4\,,
\end{align}
where $\alpha$ is static polarizability of the atom. At large distances we can
parameterize effective charge $Q(r)$ as
\begin{align}\label{eq_Qeff}
  Q(r) &= \frac{N_v}{2} \exp\left(-2\sqrt{-2\veps_v}(r-r_v)\right)\,,
\end{align}
where $\veps_v$ and $N_v$ are the electron energy and the number of electrons
for the outermost atomic shell. The radius $r_v$ can be defined so that
$Q(r_v)=\tfrac{N_v}{2}$, i.e. the screening of the nucleus at the distance
$r_v$ by the outer shell is reduced by 50\%. For the neutral closed shell atom
$r_v \sim 1$ and $N_v=4l_v+2$. The classical turning point for the positron
$r_t$ is given by the equation: $U(r_t)=\veps_p\ll 1$, or neglecting positron
energy $\veps_p$:
\begin{align}\label{eq_rt}
  r_t &\approx r_v + \frac{1}{2|2\veps_v|^{1/2}}\,
  \ln\frac{N_v r_v^3}{\alpha}
  \approx r_v + \frac{1}{2|2\veps_v|^{1/2}}\,,
\end{align}
where we approximated the logarithm by unity. We see that $r_t$ is only about
an atomic unit larger than the outer shell radius $r_v$. This is much smaller
than the typical tunneling distance in the dc electric field, making positron
annihilation much more sensitive to the inner shell contributions. On the
other hand, the distance $r_t$ is much larger than the inner shell radii, so
annihilation on the inner electrons should depend on the asymptotic behaviour
of the inner orbitals in the classically forbidden region. The estimate
\eqref{eq_rt} for the positron turning point $r_t$ is close to $r_v$ where
\Eref{eq_gen} is not applicable, but \Eref{eq_fact5a} should hold.

\begin{figure}[tbh]
 \includegraphics[scale=0.95]{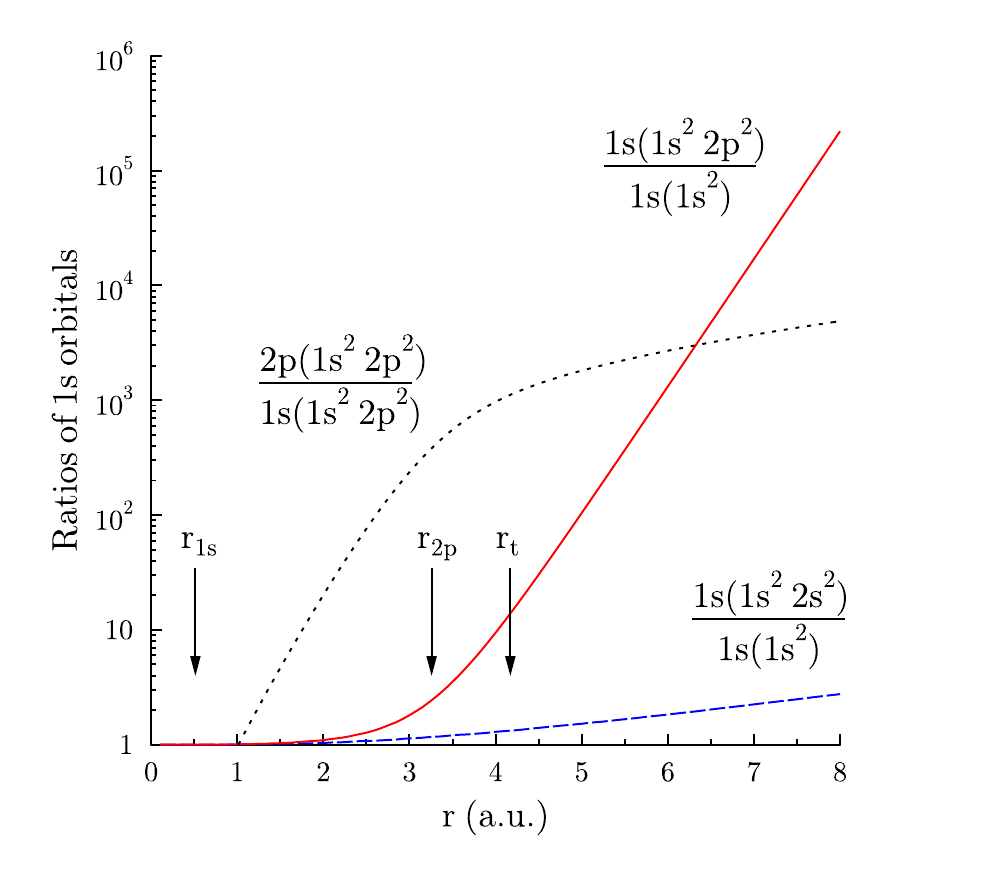}
\caption{Ratios of the $1s$ and $2p$ orbitals for different configurations of
Be. Vertical arrows show rms radii of $1s$ and $2p_{1/2}$ orbitals and the
positron turning point $r_t$.}
 \label{fig_Be_1s}
\end{figure}

As a first example we considered Be atom in different configurations and
compared asymptotics of the $1s$ orbital. We used relativistic
Hartree-Fock-Dirac code \cite{BDT77} for three closed shell configurations:
$1s^2$, $1s^2 2s^2$, and $1s^2 2p_{1/2}^2$. \fref{fig_Be_1s} shows the ratio
of the $1s$ orbitals for all three cases. One can see that for the
configurations  $1s^2$ and $1s^2 2s^2$ the long distance behaviour is similar.
The small difference is caused by a 20\% change in the $1s$ energy. We
conclude that here numerical results agree with \Eref{eq_spec}. The $1s$
energy for the configuration $1s^2 2p_{1/2}^2$ lies between the values for two
other configurations, but the asymptotics is absolutely different in agreement
with Eq.\ (\ref{eq_HF}). This is in consent with the statement that
\Eref{eq_HF} applies for the systems with occupied shells with $l\neq 0$
\cite{HMS69}. \fref{fig_Be_1s} also shows the ratio of the $2p_{1/2}$ and $1s$
orbitals for the configuration $1s^2 2p_{1/2}^2$. We see that this ratio first
grows exponentially, but then stabilizes at $r\gtrsim 3$ in agreement with
\eqref{eq_fact5a}.

Figure \ref{fig_Be_1s} shows that exchange interaction start to determine
behaviour of the $1s$ wave function at the distances $r\lesssim r_{2p}$, i.e.
near the main maximum of the $2p$ orbital. At such intermediate distances,
which lie far behind the classical turning point for the inner electron but in
the localization domain of the outer electrons, the exchange interaction is
suppressed only by a power of the radius \cite{Fla09c,Fla09a}.

For the Be atom in $1s^2 2p_{1/2}^2$ configuration the estimate \eqref{eq_rt}
gives $r_t\approx 4.2$. Thus, the distances where annihilation can take place
are $r\sim 4$. At such distances the difference between $1s$ orbitals with and
without exchange interaction (configurations $1s^2$ and $1s^2 2p_{1/2}^2$
respectively) is about one order of magnitude. Therefore, the exchange
assisted tunneling can enhance annihilation rate (which is proportional to the
probability density) by approximately two orders of magnitude. As we see from
\fref{fig_Be_1s} the ratio of the $2p$ and $1s$ orbitals at these distances is
on the order of $10^2$. Therefore, for the Be atom in the configuration $1s^2
2p^2$ we can expect the $1s$ contribution to the annihilation rate to be on
the order of $10^{-4}$, instead of $10^{-6}$ without exchange assisted
tunneling.

\begin{figure}[tbh]
 \includegraphics[scale=0.95]{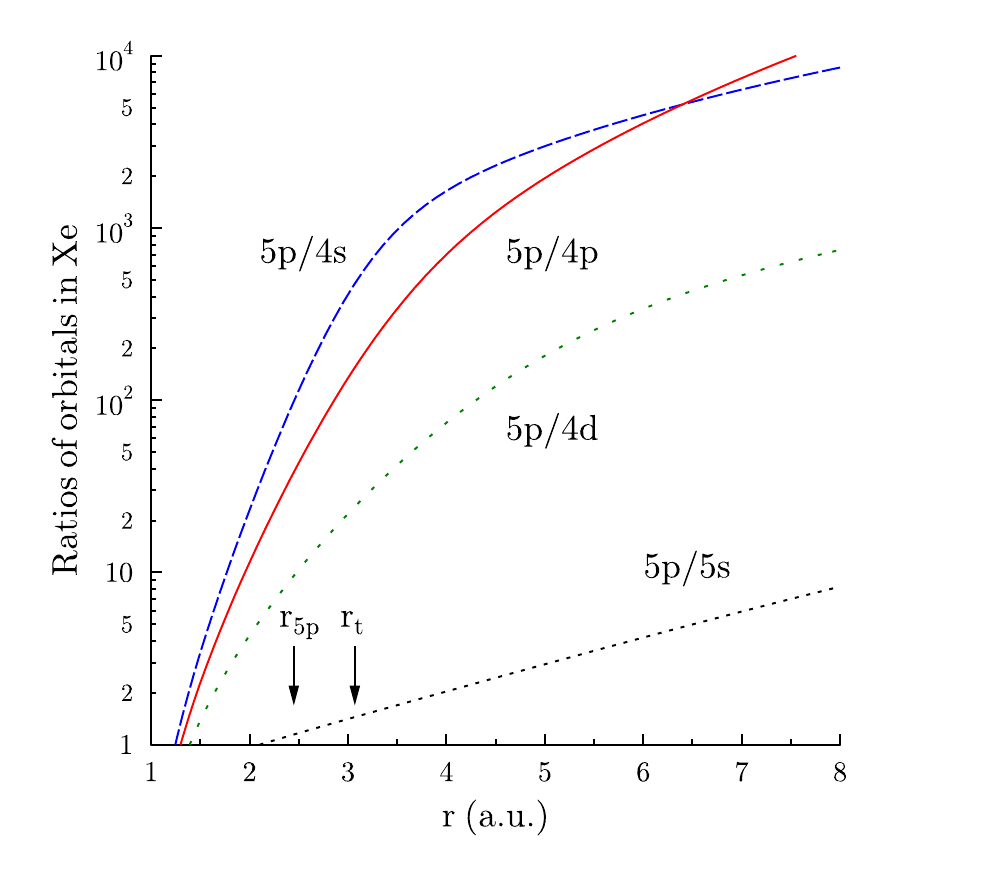}
\caption{Ratios of the $5p_{3/2}$, $5s_{1/2}$, $4d_{5/2}$, $4p_{3/2}$, and
$4s_{1/2}$ orbitals in Xe atom. Vertical arrows show rms radius of $5p_{3/2}$
orbital and the positron turning point $r_t$.}
 \label{fig_Xe}
\end{figure}

Let us consider now a more realistic example. The inner shell contributions to
the annihilation $\gamma$ line was studied by \citet{IGGS97} for several noble
gases. For Xe the total probability of the annihilation on the inner shells
$4d$, $4p$, and $4s$ was found to be 2.4\%. On \fref{fig_Xe} we plot the
ratios of the orbitals for Xe at the distances from 1 to 8 Bohr radii, where
these orbitals do not oscillate. For Xe the estimate \eqref{eq_rt} for the
classical turning point gives $r_t\approx 3.1$. At such distances the ratio of
the $5p$ and $4d$ orbitals is close to 10. Taking into account that $d$ shell
has 10 electrons while $p$ shell has only 6, we conclude that 2\% contribution
of the $4d$ shell agrees with our simple consideration. The rms radius for the
$4d$ orbital in Xe is 0.95 a.u., so such a large contribution is due to the
asymptotics \eqref{eq_fact5a}.

Correlations play very important role in the process of the positron
annihilation. They lead to the huge enhancement of the cross section. The
electron core polarization leads to the attractive potential for the positron
\eqref{eq_U} increasing the annihilation rate. Usually this effect has the
size of a typical correlation correction for many electron atoms. The dominant
correlation corrections come from the positron-electron correlations
\cite{DFKM93}. This type of correlations can lead to the virtual formation of
the positronium and neutralization of the positron charge. As a result, the
positron can penetrate deeper inside the atom. As we saw above, the classical
turning point for the positron $r_t$ lies in the region where the electron
density of the atom is exponentially decreasing, so even small correlation
corrections to $r_t$ can drastically increase the annihilation rate. However,
it was shown, by \citet{GG12} that these correlations equally affect outer and
inner shell contributions and only slightly change the annihilation line
shape. Note that this result indirectly confirms that the densities of the
inner and outer electron in the annihilation region are almost proportional to
each other, as suggested by \Eref{eq_fact5a}.

In this section we considered direct annihilation on the inner shell electron.
We argued that this process is enhanced by the exchange interaction between
inner and outer shells. The inner shell hole can be also formed in the
annihilation on the outer shell followed by ionization of the inner shell
caused by the changed atomic potential. In the lowest order of MBPT this
process is described by the Coulomb interaction between the inner and the
outer shell holes. This Coulomb integral is the same as the exchange integral
considered above with substitution of the one final orbital by the orbital in
the continuum. Such processes were considered by \citet{DuGr06}.

\section*{Conclusions}

In this paper we argue that the exchange assisted tunneling of the electrons
from the inner atomic shells is not an artifact of the Hartree-Fock
approximation, but an observable effect, at least for the intermediate
distances $r\gtrsim 1$. In particular, it can be observed in the annihilation
of the slow positrons on the many electron atoms. The annihilation on the
inner shell electrons forms the shoulders of the experimentally observed
$\gamma$ lines for the noble gases \cite{IGGS97}. It can be also directly
detected in the coincidence experiments where $\gamma$ quanta are registered
simultaneously with the Auger electrons \cite{EGKK02}. This process takes
place at the distances, comparable to the size of the outermost atomic shell.
Because of that it is much less suppressed than the inner shell dc field
ionization, which may be difficult to observe. On the other hand, without the
exchange assisted tunneling of the inner electrons to this region, the inner
shell annihilation would be unobservably small. At present we do not see any
realistic experiment to test the inner electron asymptotics at the large
distances $r\gg 1$ in atomic physics. However such asymptotics can be
important in the condensed matter physics \cite{Fla09c}.

\acknowledgments We are grateful to M.\ Yu.\ Kuchiev, V.\ A.\ Dzuba, and D. A.
Nevskii for helpful discussions. This work is partly supported by the Russian
Foundation for Basic Research Grants No.\ 11-02-00943.

\appendix

\section{Double well potential}
\label{sec_DW}

\begin{figure}[tbh]
 \includegraphics[scale=1.0]{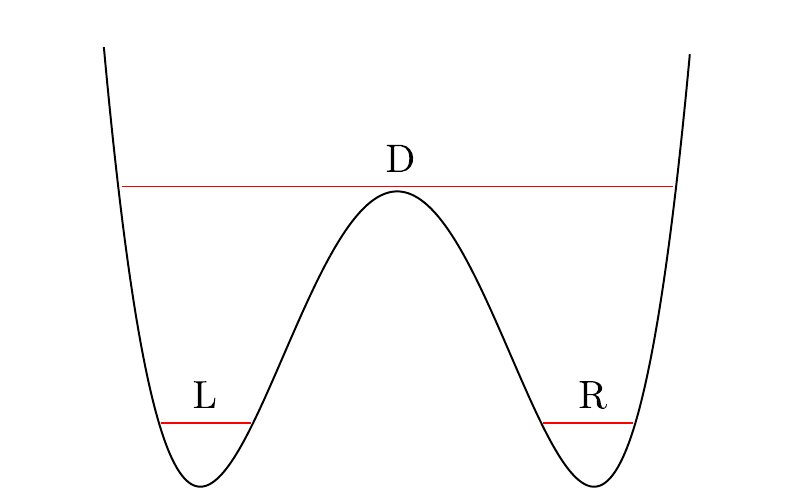}
\caption{Double well potential with two localized and one delocalized states.}
 \label{fig_dbl_well}
\end{figure}

Let us consider a double well potential. We assume that there are two
localized states L and R and one delocalized state D with energies $\veps_L$,
$\veps_R$, and $\veps_D$ (see \fref{fig_dbl_well}). If
$\veps_L=\veps_R=\veps_{LR}$, then the true eigenstates are $\psi_\pm
=(\psi_L\pm\psi_R)/\sqrt{2}$ and there is exponentially small energy splitting
between them:
 \begin{align}\label{dw1}
   &\veps_\pm=\veps_{LR}\mp \delta\,, &\delta\sim
   \langle\psi_L|\psi_R\rangle\,.
 \end{align}
The one-particle spectrum of this system is $\veps_+$, $\veps_-$, and
$\veps_D$.

If we have two non-interacting electrons in this potential in a triplet state,
the spectrum still includes three levels:
 \begin{align}\label{dw2}
   &E_a=2\veps_{LR}\,,
   &E_{b,c} =\veps_{LR}+\veps_D\mp \delta 
   \,.
 \end{align}
If we switch on interaction $1/|x_1-x_2|$ between the electrons, we get
additional splitting between levels $E_b$ and $E_c$ from the exchange
integral:
 \begin{align}\label{dw3}
   &E_c-E_b =2(\delta + \Delta)
   \,,
   \\
   &\Delta =\left\langle \psi_L(x_1)\psi_D(x_2)\left|\tfrac{1}{|x_1-x_2|}
   \right| \psi_D(x_1)\psi_R(x_2)\right\rangle
   \nonumber \\
   \label{dw4}
   &\quad\approx -\frac{2}{x_{LR}^3}
   \langle \psi_L|x| \psi_D\rangle
   \langle \psi_D|x| \psi_R\rangle
   \,.
 \end{align}
Here we left the first non-zero term of the multipolar expansion ($x_{LR}$ is
the distance between the minima of the potential).

We see that for the non-interacting electrons the splitting is caused by the
tunneling through the barrier and is of the order of the overlap integral
$\langle\psi_L|\psi_R\rangle$, which exponentially depends on the distance
$x_{LR}$. The exchange interaction with the delocalized electron results in
the splitting which depends on this distance as $x_{LR}^{-3}$. The dipole
matrix elements in \eqref{dw4} are not suppressed because the state D has
large overlaps with both localizes states. We conclude that the exchange
interaction can lead to the long range interaction between localized states.

\section{$Z^{-1}$ expansion for the He-like ion}
\label{sec_He}

The system \eqref{ap78} can be used to form the $\tfrac{1}{Z}$ expansion of
the eigenfunction of the highly charged ion. Let us consider He-like ion with
$Z\gg 1$. Making substitution $\xi_i=Zr_i$ we write Hamiltonian
\eqref{eq_fact2} as:
 \begin{multline}\label{He1}
 H^{(2)}\equiv Z^2 H^{(2)}_\xi
 \\
 =Z^2\left(-\frac12\Delta_{\xi_1}-\frac{1}{\xi_1}
 -\frac12\Delta_{\xi_2}-\frac{1}{\xi_2}
 +\frac{1}{Z}\frac{1}{\xi_{12}}
 \right)\,.
 \end{multline}
In the new variables we have
 \begin{align}\label{He2}
 &H^{(2)}_\xi\Psi^{(2)} = {\cal E}^{(2)}\Psi^{(2)}\,,
 &{\cal E}^{(2)}\equiv Z^{-2}{E}^{(2)}\,.
 \end{align}
We present the solution in the form \eqref{ap45} and need to solve system
\eqref{ap78} where
 \begin{align}
 \label{He3}
 W_{i,k}(1)
 = \frac{1}{Z}\left\langle\phi_{n_i,l_i}(2)\left|
 \frac{1}{\xi_{1,2}}\right|\phi_{n_k,l_k}(2)\right\rangle
 \,.
 \end{align}
Here the functions $\phi_{n_i,l_i}$ are hydrogenic. Explicit smallness in
\eqref{He3} and, therefore, in the right hand side of \Eref{ap7} allows
expansion in $Z^{-1}$.

Let us consider the triplet state $1s2p_{m=0}$. The zero-order orbital
functions have the form:
 \begin{align}
 \label{He4}
 \Psi_0^{(2)} &= \frac{1}{\sqrt{2}}
 \big(\phi_{2p_0}(1)\phi_{1s}(2)-\phi_{1s}(1)\phi_{2p_0}(2)\big)\,,
 \\
 \label{He5}
 \phi_{1s}(\xi) &= \frac{2}{\sqrt{4\pi}} \mathrm{e}^{-\xi}\,,
 \\
 \label{He6}
 \phi_{2p_0}(\xi) &= \frac{\cos\theta}{4\sqrt{2\pi}}
 \xi\,\mathrm{e}^{-\xi/2}\,.
 \end{align}

Note that already in the first order in $Z^{-1}$ we have infinite number of
channels. However, these channels correspond to the different final states of
the ion and can be considered independently. We are interested only in the
first order corrections to the orbitals $f_1$ and $f_2$, which correspond to
the ion either in the $1s$, or in the $2p$ state:
 \begin{align}
 \label{He7}
 f_1 &=\phi_{2p_0} + f^{(1)}_{2p_0}\,,
 \\
 \label{He8}
 f_2 &=\phi_{1s} + f^{(1)}_{1s} + f^{(1)}_{d_0}\,.
 \end{align}
Note that angular and parity selection rules allow mixing of $s$- and
$d$-waves for the amplitude $f_2$. However, we will be interested only in the
correction to the $s$-wave $f^{(1)}_{1s}$.

For the orbitals \eqref{He5} and \eqref{He6} the functions $W_{i,k}$ can be
found analytically using \Eref{He3}:
 \begin{align}
 \label{He9}
 W_{1,1}
 &= \frac{1}{\sqrt{2}Z\xi}\Big(
 1-\mathrm{e}^{-2\xi}(\xi+1)
 \Big)\,,
 \\
 \label{He10}
 W_{1,2}
 &= -\frac{2\cos\theta}{243Z\xi^2}\Big(
 64-\mathrm{e}^{-3\xi/2}(27\xi^3
 \nonumber\\
 &\qquad
 +72\xi^2+96\xi+64)
 \Big)\,,
 \\
 \label{He11}
 \tilde{W}_{2,1}
 &= \frac{2}{729Z\xi^2\cos\theta}\Big(
 64-\mathrm{e}^{-3\xi/2}(27\xi^3
 \nonumber\\
 &\qquad
 +72\xi^2+96\xi+64)
 \Big)\,,
 \\
 \label{He12}
 W_{2,2}
 &= -\frac{1}{\sqrt{2}Z\xi}\Big(
 1-\frac{1}{24}\mathrm{e}^{-\xi}(\xi^3
 \nonumber\\
 &\qquad
 +6\xi^2+18\xi+24)
 \Big)\,.
 \end{align}
In \Eref{He11} we left only the term for which the product
$\phi_{2p_0}\tilde{W}_{2,1}$ has no angular dependence and, therefore,
contributes to the $s$-wave part of the amplitude \eqref{He8}.

The first order corrections $f^{(1)}_{2p_0}$ and $f^{(1)}_{1s}$ satisfy
equations:
 \begin{multline}\label{He13}
 \left(-\frac12\Delta_{\xi}-\frac{1}{\xi} +\frac{1}{8}
 \right)
 f^{(1)}_{2p_0}
 \\
 =-\phi_{2p_0}W_{1,1}-\phi_{1s}W_{1,2}
 \,,
 \end{multline}
 \begin{multline}\label{He14}
 \left(-\frac12\Delta_{\xi}-\frac{1}{\xi} +\frac{1}{2}
 \right)
 f^{(1)}_{1s}
 \\
 =-\phi_{2p_0}\tilde{W}_{2,1}-\phi_{1s}W_{2,2}
 \,.
 \end{multline}
These equations are valid at all distances. We see that the right hand sides
of both of them have three different exponents, $-\xi/2$, $-\xi$, and
$-5\xi/2$. The long-range asymptotics depends on the weakest exponent
$-\xi/2$, which corresponds to the binding energy of the electron $2p$. We
conclude that the first term of the $Z^{-1}$ expansion has asymptotics in
agreement with the general case discussed above.


\begin{thebibliography}{20}
\expandafter\ifx\csname natexlab\endcsname\relax\def\natexlab#1{#1}\fi
\expandafter\ifx\csname bibnamefont\endcsname\relax
  \def\bibnamefont#1{#1}\fi
\expandafter\ifx\csname bibfnamefont\endcsname\relax
  \def\bibfnamefont#1{#1}\fi
\expandafter\ifx\csname citenamefont\endcsname\relax
  \def\citenamefont#1{#1}\fi
\expandafter\ifx\csname url\endcsname\relax
  \def\url#1{\texttt{#1}}\fi
\expandafter\ifx\csname urlprefix\endcsname\relax\def\urlprefix{URL }\fi
\providecommand{\bibinfo}[2]{#2} \providecommand{\eprint}[2][]{\url{#2}}

\bibitem[{\citenamefont{{Handy} et~al.}(1969)\citenamefont{{Handy}, {Marron},
  and {Silverstone}}}]{HMS69}
\bibinfo{author}{\bibfnamefont{N.~C.} \bibnamefont{{Handy}}},
  \bibinfo{author}{\bibfnamefont{M.~T.} \bibnamefont{{Marron}}},
  \bibnamefont{and} \bibinfo{author}{\bibfnamefont{H.~J.}
  \bibnamefont{{Silverstone}}}, \bibinfo{journal}{Physical Review}
  \textbf{\bibinfo{volume}{180}}, \bibinfo{pages}{45} (\bibinfo{year}{1969}).

\bibitem[{\citenamefont{Dzuba et~al.}(1982)\citenamefont{Dzuba, Flambaum, and
  Silvestrov}}]{DFS82}
\bibinfo{author}{\bibfnamefont{V.~A.} \bibnamefont{Dzuba}},
  \bibinfo{author}{\bibfnamefont{V.~V.} \bibnamefont{Flambaum}},
  \bibnamefont{and} \bibinfo{author}{\bibfnamefont{P.~G.}
  \bibnamefont{Silvestrov}}, \bibinfo{journal}{J. Phys. B}
  \textbf{\bibinfo{volume}{15}}, \bibinfo{pages}{L575} (\bibinfo{year}{1982}).

\bibitem[{\citenamefont{{Flambaum}}(2009{\natexlab{a}})}]{Fla09c}
\bibinfo{author}{\bibfnamefont{V.~V.} \bibnamefont{{Flambaum}}},
  \bibinfo{journal}{Phys. Rev. A} \textbf{\bibinfo{volume}{79}},
  \bibinfo{eid}{042505} (\bibinfo{year}{2009}{\natexlab{a}}),
  \eprint{arXiv:0809.2847}.

\bibitem[{\citenamefont{{Morrell} et~al.}(1975)\citenamefont{{Morrell}, {Parr},
  and {Levy}}}]{MPL75}
\bibinfo{author}{\bibfnamefont{M.~M.} \bibnamefont{{Morrell}}},
  \bibinfo{author}{\bibfnamefont{R.~G.} \bibnamefont{{Parr}}},
  \bibnamefont{and} \bibinfo{author}{\bibfnamefont{M.}~\bibnamefont{{Levy}}},
  \bibinfo{journal}{J. Chem. Phys.} \textbf{\bibinfo{volume}{62}},
  \bibinfo{pages}{549} (\bibinfo{year}{1975}).

\bibitem[{\citenamefont{{Katriel} and {Davidson}}(1980)}]{KaDa80}
\bibinfo{author}{\bibfnamefont{J.}~\bibnamefont{{Katriel}}} \bibnamefont{and}
  \bibinfo{author}{\bibfnamefont{E.~R.} \bibnamefont{{Davidson}}},
  \bibinfo{journal}{Proceedings of the National Academy of Science}
  \textbf{\bibinfo{volume}{77}}, \bibinfo{pages}{4403} (\bibinfo{year}{1980}).

\bibitem[{\citenamefont{{Flambaum}}(2009{\natexlab{b}})}]{Fla09b}
\bibinfo{author}{\bibfnamefont{V.~V.} \bibnamefont{{Flambaum}}},
  \bibinfo{journal}{Phys. Rev. A} \textbf{\bibinfo{volume}{80}},
  \bibinfo{eid}{055401} (\bibinfo{year}{2009}{\natexlab{b}}),
  \eprint{arXiv:0909.1868}.

\bibitem[{\citenamefont{{Ruderman} and {Kittel}}(1954)}]{RuKi54}
\bibinfo{author}{\bibfnamefont{M.~A.} \bibnamefont{{Ruderman}}}
  \bibnamefont{and} \bibinfo{author}{\bibfnamefont{C.}~\bibnamefont{{Kittel}}},
  \bibinfo{journal}{Physical Review} \textbf{\bibinfo{volume}{96}},
  \bibinfo{pages}{99} (\bibinfo{year}{1954}).

\bibitem[{\citenamefont{{Kasuya}}(1956)}]{Kas56}
\bibinfo{author}{\bibfnamefont{T.}~\bibnamefont{{Kasuya}}},
  \bibinfo{journal}{Progress of Theoretical Physics}
  \textbf{\bibinfo{volume}{16}}, \bibinfo{pages}{45} (\bibinfo{year}{1956}).

\bibitem[{\citenamefont{{Yosida}}(1957)}]{Yos57}
\bibinfo{author}{\bibfnamefont{K.}~\bibnamefont{{Yosida}}},
  \bibinfo{journal}{Physical Review} \textbf{\bibinfo{volume}{106}},
  \bibinfo{pages}{893} (\bibinfo{year}{1957}).

\bibitem[{\citenamefont{{Fisher} et~al.}(1998)\citenamefont{{Fisher}, {Maron},
  and {Pitaevskii}}}]{FMP98}
\bibinfo{author}{\bibfnamefont{D.}~\bibnamefont{{Fisher}}},
  \bibinfo{author}{\bibfnamefont{Y.}~\bibnamefont{{Maron}}}, \bibnamefont{and}
  \bibinfo{author}{\bibfnamefont{L.~P.} \bibnamefont{{Pitaevskii}}},
  \bibinfo{journal}{Phys. Rev. A} \textbf{\bibinfo{volume}{58}},
  \bibinfo{pages}{2214} (\bibinfo{year}{1998}).

\bibitem[{\citenamefont{{Flambaum}}(2009{\natexlab{c}})}]{Fla09a}
\bibinfo{author}{\bibfnamefont{V.~V.} \bibnamefont{{Flambaum}}},
  \bibinfo{journal}{EPL (Europhysics Letters)} \textbf{\bibinfo{volume}{88}},
  \bibinfo{pages}{60009} (\bibinfo{year}{2009}{\natexlab{c}}),
  \eprint{arXiv:0910.1155}.

\bibitem[{\citenamefont{{Amusia}}(2009)}]{Amu09}
\bibinfo{author}{\bibfnamefont{M.~Y.} \bibnamefont{{Amusia}}},
  \bibinfo{journal}{Soviet Journal of Experimental and Theoretical Physics
  Letters} \textbf{\bibinfo{volume}{90}}, \bibinfo{pages}{161}
  (\bibinfo{year}{2009}), \eprint{arXiv:0904.4395}.

\bibitem[{\citenamefont{{Iwata} et~al.}(1997)\citenamefont{{Iwata}, {Gribakin},
  {Greaves}, and {Surko}}}]{IGGS97}
\bibinfo{author}{\bibfnamefont{K.}~\bibnamefont{{Iwata}}},
  \bibinfo{author}{\bibfnamefont{G.~F.} \bibnamefont{{Gribakin}}},
  \bibinfo{author}{\bibfnamefont{R.~G.} \bibnamefont{{Greaves}}},
  \bibnamefont{and} \bibinfo{author}{\bibfnamefont{C.~M.}
  \bibnamefont{{Surko}}}, \bibinfo{journal}{Phys. Rev. Lett.}
  \textbf{\bibinfo{volume}{79}}, \bibinfo{pages}{39} (\bibinfo{year}{1997}).

\bibitem[{\citenamefont{{Eshed} et~al.}(2002)\citenamefont{{Eshed},
  {Goktepeli}, {Koymen}, {Kim}, {Chen}, {O'Kelly}, {Sterne}, and
  {Weiss}}}]{EGKK02}
\bibinfo{author}{\bibfnamefont{A.}~\bibnamefont{{Eshed}}},
  \bibinfo{author}{\bibfnamefont{S.}~\bibnamefont{{Goktepeli}}},
  \bibinfo{author}{\bibfnamefont{A.~R.} \bibnamefont{{Koymen}}},
  \bibinfo{author}{\bibfnamefont{S.}~\bibnamefont{{Kim}}},
  \bibinfo{author}{\bibfnamefont{W.~C.} \bibnamefont{{Chen}}},
  \bibinfo{author}{\bibfnamefont{D.~J.} \bibnamefont{{O'Kelly}}},
  \bibinfo{author}{\bibfnamefont{P.~A.} \bibnamefont{{Sterne}}},
  \bibnamefont{and} \bibinfo{author}{\bibfnamefont{A.~H.}
  \bibnamefont{{Weiss}}}, \bibinfo{journal}{Phys. Rev. Lett.}
  \textbf{\bibinfo{volume}{89}}, \bibinfo{eid}{075503} (\bibinfo{year}{2002}).

\bibitem[{\citenamefont{{Dunlop} and {Gribakin}}(2006)}]{DuGr06}
\bibinfo{author}{\bibfnamefont{L.~J.~M.} \bibnamefont{{Dunlop}}}
  \bibnamefont{and} \bibinfo{author}{\bibfnamefont{G.~F.}
  \bibnamefont{{Gribakin}}}, \bibinfo{journal}{J. Phys. B}
  \textbf{\bibinfo{volume}{39}}, \bibinfo{pages}{1647} (\bibinfo{year}{2006}),
  \eprint{arXiv:physics/0512175}.

\bibitem[{\citenamefont{Wang et~al.}(2010)\citenamefont{Wang, Selvam, Gribakin,
  and Surko}}]{WSGS10}
\bibinfo{author}{\bibfnamefont{F.}~\bibnamefont{Wang}},
  \bibinfo{author}{\bibfnamefont{L.}~\bibnamefont{Selvam}},
  \bibinfo{author}{\bibfnamefont{G.~F.} \bibnamefont{Gribakin}},
  \bibnamefont{and} \bibinfo{author}{\bibfnamefont{C.~M.} \bibnamefont{Surko}},
  \bibinfo{journal}{J. Phys. B} \textbf{\bibinfo{volume}{43}},
  \bibinfo{pages}{165207} (\bibinfo{year}{2010}).

\bibitem[{\citenamefont{{Surko} et~al.}(2005)\citenamefont{{Surko}, {Gribakin},
  and {Buckman}}}]{SGB05}
\bibinfo{author}{\bibfnamefont{C.~M.} \bibnamefont{{Surko}}},
  \bibinfo{author}{\bibfnamefont{G.~F.} \bibnamefont{{Gribakin}}},
  \bibnamefont{and} \bibinfo{author}{\bibfnamefont{S.~J.}
  \bibnamefont{{Buckman}}}, \bibinfo{journal}{J. Phys. B}
  \textbf{\bibinfo{volume}{38}}, \bibinfo{pages}{57} (\bibinfo{year}{2005}).

\bibitem[{\citenamefont{Dzuba et~al.}(1993)\citenamefont{Dzuba, Flambaum, King,
  Miller, and Sushkov}}]{DFKM93}
\bibinfo{author}{\bibfnamefont{V.~A.} \bibnamefont{Dzuba}},
  \bibinfo{author}{\bibfnamefont{V.~V.} \bibnamefont{Flambaum}},
  \bibinfo{author}{\bibfnamefont{W.~A.} \bibnamefont{King}},
  \bibinfo{author}{\bibfnamefont{B.~N.} \bibnamefont{Miller}},
  \bibnamefont{and} \bibinfo{author}{\bibfnamefont{O.~P.}
  \bibnamefont{Sushkov}}, \bibinfo{journal}{Physica Scripta}
  \textbf{\bibinfo{volume}{T46}}, \bibinfo{pages}{248} (\bibinfo{year}{1993}).

\bibitem[{\citenamefont{Bratsev et~al.}(1977)\citenamefont{Bratsev, Deyneka,
  and Tupitsyn}}]{BDT77}
\bibinfo{author}{\bibfnamefont{V.~F.} \bibnamefont{Bratsev}},
  \bibinfo{author}{\bibfnamefont{G.~B.} \bibnamefont{Deyneka}},
  \bibnamefont{and} \bibinfo{author}{\bibfnamefont{I.~I.}
  \bibnamefont{Tupitsyn}}, \bibinfo{journal}{Bull. Acad. Sci. USSR, Phys. Ser.}
  \textbf{\bibinfo{volume}{41}}, \bibinfo{pages}{173} (\bibinfo{year}{1977}).

\bibitem[{\citenamefont{{Green} and {Gribakin}}(2012)}]{GG12}
\bibinfo{author}{\bibfnamefont{D.~G.} \bibnamefont{{Green}}} \bibnamefont{and}
  \bibinfo{author}{\bibfnamefont{G.~F.} \bibnamefont{{Gribakin}}},
  \bibinfo{journal}{Journal of Physics Conference Series}
  \textbf{\bibinfo{volume}{388}}, \bibinfo{pages}{072018}
  (\bibinfo{year}{2012}).

\end{thebibliography}

\end{document}